\pdfoutput=1
\documentclass[aps,twocolumn, floatfix, prl]{revtex4}
\usepackage{graphicx}
\DeclareGraphicsExtensions{.pdf}
\usepackage{amsmath,amssymb,bbold,bm}
\usepackage{float}
\usepackage{epstopdf}

\newcommand{\bk}{{\bm k}}

\newcommand{\br}{{\bm r}}
\newcommand{\bR}{{\bm R}}
\newcommand{\bB}{{\bm B}}

\newcommand{\bS}{{\bm S}}

\newcommand{\bs}{{\bm s}}

\newcommand{\cH}{{\cal H}}
\newcommand{\cG}{{\cal G}}

\begin{document}

\title{Self-organized topological state with Majorana fermions}

\author{M.M. Vazifeh}
\author{M. Franz}
\affiliation{Department of Physics and Astronomy, University of
British Columbia, Vancouver, BC, Canada V6T 1Z1}

\begin{abstract}
Most physical systems known to date tend to resist entering the topological phase and must be fine tuned to reach the latter. Here we introduce a system in which a key dynamical parameter adjusts itself in response to the changing external conditions so that the ground state naturally favors the topological phase.  The system  consists of a quantum wire formed of individual magnetic atoms placed on the surface of an ordinary $s$-wave superconductor. It realizes the Kitaev paradigm of topological superconductivity when the wavevector characterizing the emergent spin helix dynamically self-tunes to support the topological phase. We call this phenomenon  self-organized topological state. 
\end{abstract}

\date{\today}

\maketitle

Topological phases, quite generally, are difficult to come by. They either occur under rather extreme conditions (e.g. the quantum Hall liquids  \cite{prange1}, which require high sample purity, strong magnetic fields and low temperatures) or demand fine tuning of system parameters, as in the majority of known topological insulators \cite{moore_rev,hasan_rev,qi_rev}. Many perfectly sensible topological phases, such as the Weyl semimetals \cite{weyl4} and topological superconductors \cite{qi_rev,kallin1}, remain experimentally undiscovered. 

The paucity of easily accessible, stable topological materials has been in a large part responsible for the relatively slow progress towards the adoption of topological phases in the mainstream technological applications. A question that naturally arises is the following: Is there a fundamental principle behind this ``topological resistance''?
Although unable to give a general answer to this question we provide in this Letter a specific counterexample to this conjectured phenomenon of topological resistance. We consider a simple model system which, as we demonstrate, {\em wants} to be topological in a precisely defined sense. The key to this ``topofilia'' is the existence in the system of a dynamical parameter that adjusts itself in response to changing external conditions so that the system self-tunes into the topological phase. 
\begin{figure}[b]
\includegraphics[width = 7.6cm]{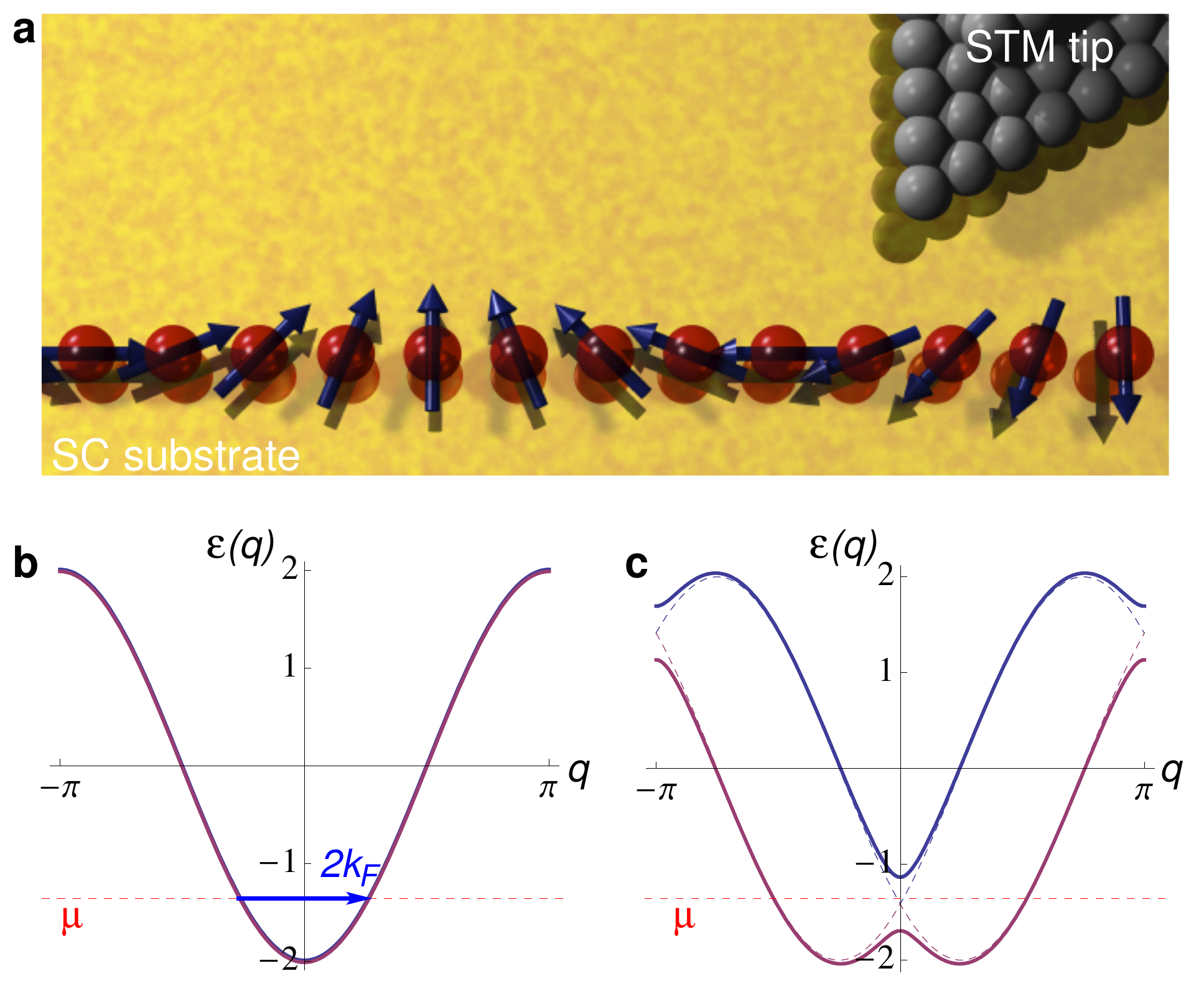}
\caption{{\bf Chain of magnetic atoms on a superconducting substrate.} a) Schematic depiction of the system with the red spheres representing the adatoms and blue arrows showing their magnetic moments arranged in a spiral. b) Two spin-degenerate branches of the normal-state spectrum of the system in the absence of magnetic moments modeled by the nearest-neighbour tight-binding model Eq.\ (\ref{hspi1}). c) With the magnetic moments the two branches shift in momentum by $\pm G$  and the gap $JS$ opens at $q=0,\pi$. Dashed lines show the shifted spectral branches indicated in panel (b) with no gap for comparison. 
}\label{fig1}
\end{figure}
The specific model system we consider is depicted in Fig.\ \ref{fig1}a and consists of a chain of magnetic atoms, such as Co, Mn or Fe, deposited on the atomically flat surface of an ordinary $s$-wave superconductor, as described in a recent experimental study \cite{nadj-perge1}. We note that scanning tunnelling microscopy (STM) techniques now enable fairly routine assembly of such and even much more complicated nanostructures \cite{manoharan1,manoharan2}.

 It has been pointed out previously \cite{choy1,martin1,nadj-perge2} that if the magnetic moments in the chain exhibit a spiral order then the electrons in the chain can form a 1D topological superconductor (TSC) with Majorana zero modes localized at its ends \cite{kitaev1}. For a given chemical potential $\mu$, however, the spiral must have the correct pitch in order to support the topological phase. This connection is illustrated in Fig.\ \ref{fig1}b,c and will be discussed in more detail below. Exactly how the pitch of the spiral depends on the system parameters and its thermodynamic stability are two key issues that have not been previously discussed. In this Letter we show that, remarkably, under generic conditions the pitch of the spiral that minimizes the free energy of the system coincides with the one required to establish the topological phase. 

The physics behind the self-organization phenomenon outlined above is easy to understand and is similar to that leading to the spiral ordering of nuclear spins proposed to occur in 1D conducting wires \cite{loss1,loss3} and 2D electron gases \cite{loss2}. Some experimental evidence for such an ordering has been reported \cite{menzel1,scheller1}. If we for a moment neglect the superconducting order and assume a weak coupling of the adatoms to the substrate then the electrons in the chain can be thought of as forming a 1D metal. The natural wavevector for the spiral ordering in such a 1D metal is $G=2k_F$ where $k_F$ denotes its Fermi momentum. This is because the static spin susceptibility $\chi_0(q)$ of a 1D metal has a divergence at $q=2k_F$. Electron scattering off of such a magnetic spiral results in opening of a gap in the electron excitation spectrum but only for one of the two spin-degenerate bands \cite{loss1,loss3}. In the end, we are left with a single, non-degenerate Fermi crossing at $\pm 2k_F$, illustrated in Fig.\ \ref{fig1}c.  According to the Kitaev criterion \cite{kitaev1} this is exactly the condition necessary for a 1D TSC to emerge. In the following we will show that this reasoning remains valid when we include superconductivity from the outset and when we describe the chain by a tight-binding model appropriate for a discrete atomic chain.     
\begin{figure}[t]
\includegraphics[width = 6.4cm]{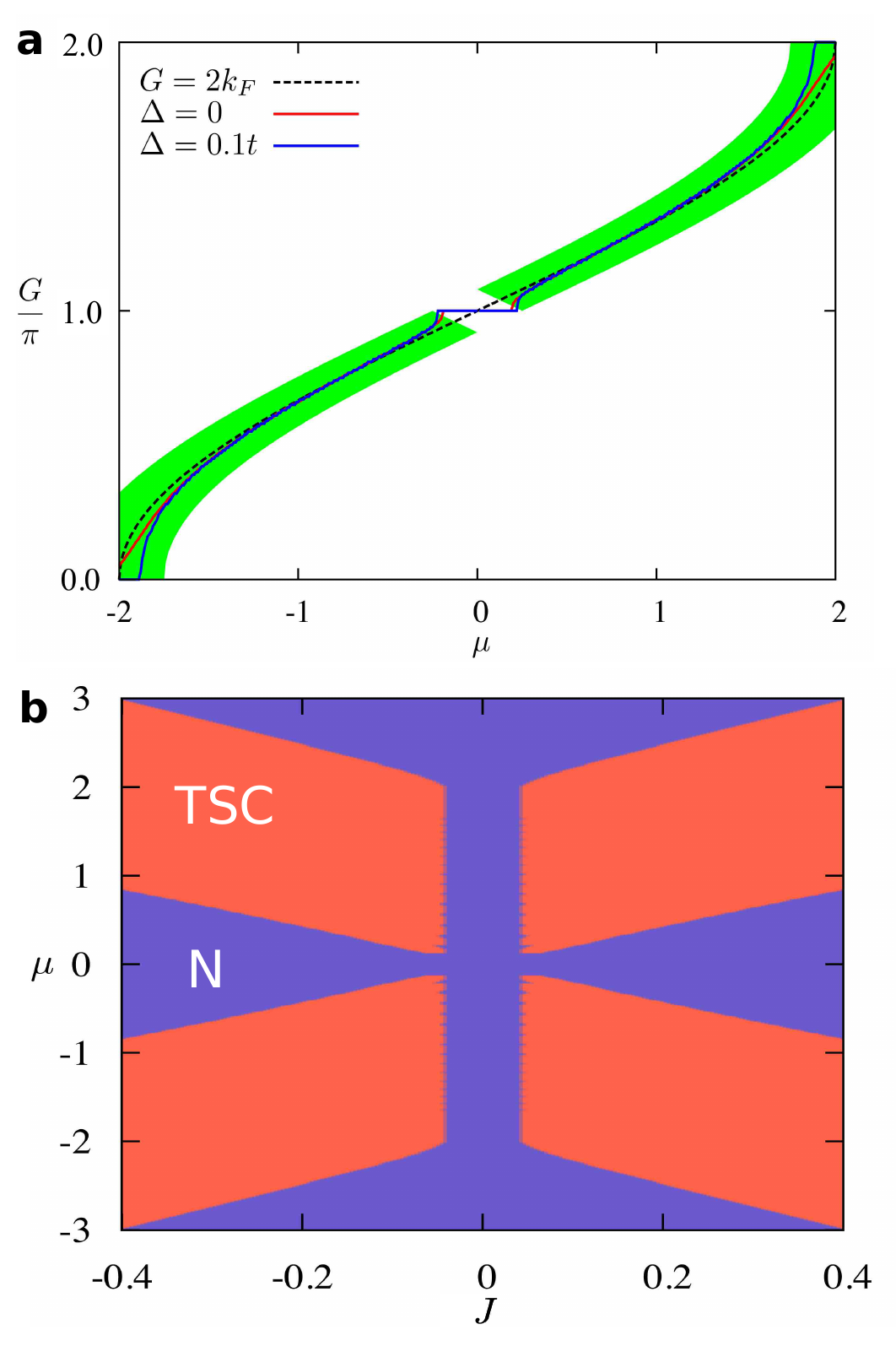}
\caption{{\bf The pitch of the spiral and the topological phase diagram.} Panel {\bf a} shows the spiral wavevector $G$ that minimizes the system ground state energy $E_g(G)$ as a function of $\mu$, the latter in units of $t$.  The parameters are $S=5/2$, $J=0.1t$ and $\Delta=0$ (red), $\Delta=0.1t$ (blue). The dashed line represents $G=2k_F$  while the green band shows the region in which $G$ must lie for the system to be topological for a given $\mu$. Panel {\bf b} shows the topological phase diagram in the $\mu$--$J$ plane, both in units of $t$, for $\Delta=0.1t$. To distinguish the two phases we have calculated the Majorana number ${\cal M}$ as defined in Ref.\ \onlinecite{kitaev1}. Topological phase (TSC) is indicated when ${\cal M}=-1$ while ${\cal M}=+1$ indicates the topologically trivial phase (N).
}\label{fig2}
\end{figure}

We begin by studying the simplest model of tight-binding electrons coupled to magnetic moments $\bS_i$ described by the Hamiltonian
\begin{equation}\label{hspi1}
{\cal H}_0=-\sum_{ij\sigma}t_{ij} c^\dagger_{i\sigma} c_{j\sigma} - \mu\sum_{i\sigma}  c^\dagger_{i\sigma} c_{i\sigma}
 +J\sum_{i}  \bS_i\cdot(c^\dagger_{i\sigma}{\bm \sigma}_{\sigma\sigma'} c_{i\sigma'})
\end{equation}
Here $c^\dagger_{j\sigma}$ creates an electron with spin $\sigma$ on site $j$,
$J$ stands for the exchange coupling constant and ${\bm \sigma}=(\sigma^x,\sigma^y,\sigma^z)$ is the vector of Pauli spin matrices.
 We assume that the substrate degrees of freedom have been integrated out, leading to a superconducting order $\Delta$ in the chain  described by
\begin{equation}\label{hspi1a}
{\cal H}={\cal H}_0+ \sum_j(\Delta c^\dagger_{j\uparrow} c^\dagger_{j\downarrow}+{\rm h.c.}).
\end{equation}
We consider a co-planar helical arrangement of atomic spins as indicated in Fig.\ \ref{fig1}a,
\begin{equation}\label{hspi2}
\bS_j=S[\cos{(Gx_j)},\sin{(Gx_j)},0]
\end{equation}
where $G$ is the corresponding wavevector and the chain is assumed to lie along the $x$-axis. We note that Hamiltonian (\ref{hspi1a}) is invariant under the simultaneous global SU(2) rotation of the electron and atomic spins so the discussion below in fact applies to any co-planar spiral.

To find the spectrum of excitations it is useful to perform a spin-dependent gauge transformation  \cite{martin1},
\begin{equation}\label{hspi3}
c_{j\uparrow}\to c_{j\uparrow} e^{{i\over 2}Gx_j}, \ \ \ \ 
c_{j\downarrow}\to c_{j\downarrow} e^{-{i\over 2}Gx_j},
\end{equation}
upon which the Hamiltonian becomes translationally invariant and can be written in the momentum space as 
\begin{eqnarray}\label{hspi4}
{\cal H}&=&\sum_q\bigl[\xi(q)c^\dagger_{q\sigma}c_{q\sigma}
+b(q)c^\dagger_{q\sigma}\sigma^z_{\sigma\sigma'}c_{q\sigma'} \\
&+& JSc^\dagger_{q\sigma}\sigma^x_{\sigma\sigma'}c_{q\sigma'}
+(\Delta c^\dagger_{q\uparrow} c^\dagger_{-q\downarrow}+{\rm h.c.})\bigr].\nonumber
\end{eqnarray}
In the above
$\xi(q)={1\over2}[\epsilon_0(q-G/2)+\epsilon_0(q+G/2)]-\mu$, and 
$b(q)={1\over2}[\epsilon_0(q-G/2)-\epsilon_0(q+G/2)]$
with $\epsilon_0(q)=-\sum_jt_{0j}e^{iqx_j}$ the normal-state dispersion in the absence of the exchange coupling.  We note that the Hamiltonian (\ref{hspi4}) is essentially a lattice version of the model semiconductor wire studied in Refs.\ \onlinecite{sau2,oreg1} with $b(q)$ playing the role of the spin-orbit coupling and $JS$ standing for the Zeeman field. Its normal state spectrum is given by
\begin{equation}\label{hspi6}
\epsilon(q)=\xi(q)\pm\sqrt{b(q)^2+J^2S^2},
\end{equation}
and is displayed in Fig.\ \ref{fig1}c  for the case of nearest-neighbour hopping with $\epsilon_0(q)=-2t\cos{q}$.

If viewed as a rigid band structure then according to the Kitaev criterion \cite{kitaev1} the chain will support topological superconductivity when there is an odd number of Fermi crossings in the right half of the Brillouin zone. This requires $\mu$ such that $|\mu\pm 2t\cos{(G/2)}|<JS$. However, in the SU(2) symmetric model under consideration, $G$ is a dynamical parameter that will assume a value that minimizes the system free energy. Taking $\bS_i$ to be classical magnetic moments and working at $T=0$ we thus proceed to minimize the ground state energy of the electrons $E_g(G)$ for a given value of $\mu$ and  $\Delta$. The result of this procedure is shown in Fig.\ \ref{fig2}a and confirms that at minimum $G\approx 2k_F$, as suggested by the general arguments advanced above. More importantly, for almost all relevant values of $\mu$ and $\Delta$  the self-consistently determined spiral pitch $G$ is precisely the one required for the formation of the topological phase. This fails only close to the half filling ($\mu =0$) where $G={\pi}$ indicates an antiferromagnetic ordering. In this case the symmetry of the band structure prohibits an odd number of Fermi crossings so the system must be in the trivial phase. Also, it is clear that no value of $G$ can bring about the TSC phase when $\mu$ lies outside of the tight-binding  band and the system is an insulator.
The resulting topological phase diagram is displayed in Fig.\ \ref{fig2}b.

These results indicate that, as we argued on general grounds, the pitch of the magnetic spiral self-tunes into the topological phase for nearly all values of the chemical potential $\mu$ for which such a tuning is possible. The emergence of Majorana zero modes at the two ends of such a topological wire \cite{kitaev1,sau2,oreg1} and their significance for the quantum information processing have been amply discussed in the recent literature  \cite{alicea_rev,beenakker_rev,stanescu_rev}. 

We now address the  adatom coupling to the substrate in greater detail. We consider a more complete Hamiltonian $\cH=\cH_0+\cH_{\rm SC} +\cH_{cd}$, where $\cH_0$ is defined in Eq.\ (\ref{hspi1}), while
\begin{equation}\label{hspi10}
{\cal H}_{\rm SC}=\sum_\bk\bigl[\xi_0(\bk)d^\dagger_{\bk\sigma}d_{\bk\sigma}
+(\Delta_0 d^\dagger_{\bk\uparrow} d^\dagger_{-\bk\downarrow}+{\rm h.c.})\bigr]
\end{equation}
describes the SC substrate with electron operators $d^\dagger_{\bk\sigma}$. The substrate is characterized by a  three-dimensional normal-state dispersion $\xi_0(\bk)=k^2/2m-\epsilon_F$ and the bulk gap amplitude $\Delta_0$. The coupling is effected through 
\begin{equation}\label{hspi11}
{\cal H}_{cd}=-r\sum_{j\sigma}(d^\dagger_{j\sigma}c_{j\sigma}+{\rm h.c.}), 
\end{equation}
where $d_{j\sigma}={1\over\sqrt{N}}\sum_\bk e^{-i\bk\cdot\bR_j}d_{\bk\sigma}$,
$N$ is the number of adatoms in the chain and $\bR_j$ denotes their positions.

We now wish to integrate out the substrate degrees of freedom and ascertain their effect on the magnetic chain. Since the Hamiltonian $\cH$ is non-interacting this procedure can be performed exactly. As outlined in the Appendix of Ref.\ \onlinecite{alicea_rev}, a  simple result obtains in the limit of the substrate  bandwidth much larger than the chain bandwidth $4t$, which we expect to generically be the case. In this limit the Green's function of the chain reads
\begin{equation}\label{hspi12}
{\cG}_{\rm eff}^{-1}(i\omega_n,q)={\cG}_{0}^{-1}(i\omega_n,q)-
\pi\rho_0r^2{i\omega_n-\tau^x\Delta_0\over\sqrt{\omega_n^2+\Delta_0^2}},
\end{equation}
where $\omega_n=(2n+1)\pi T$ is the Matsubara frequency, $\rho_0=ma^2/2\pi\hbar^2$ is the substrate normal density of states projected onto the chain (with $a$ the adatom spacing)  and
\begin{equation}\label{hspi13}
{\cG}_{0}^{-1}(i\omega_n,q)=-i\omega_n+\tau^z\left[\xi(q)+\sigma^z b(q)\right]+\sigma^x JS
\end{equation}
the bare chain Green's function. The above Green's functions are $4\times 4$ matrices in the combined spin and particle-hole (Nambu) space, the latter represented by a vector of Pauli matrices ${\bm \tau}$. In the low-frequency limit $\omega\ll \Delta_0$, relevant to the physics close to the Fermi energy, Eq.\ (\ref{hspi12}) implies two effects. First, the bare chain parameters $t$, $\mu$ and $J$ are reduced by a factor of $\Delta_0/(\Delta_0+\pi r^2\rho_0)$. Second, a SC gap $\Delta=\pi r^2\rho_0\Delta_0/(\Delta_0+\pi r^2\rho_0)$ is induced in the chain. In the limit of a weak chain-substrate coupling, $r^2\ll \Delta_0/\rho_0$, the later is seen to become $\Delta\simeq\pi r^2\rho_0$, independent of the substrate gap $\Delta_0$.

\begin{figure*}[t]
\includegraphics[width = 11.2cm]{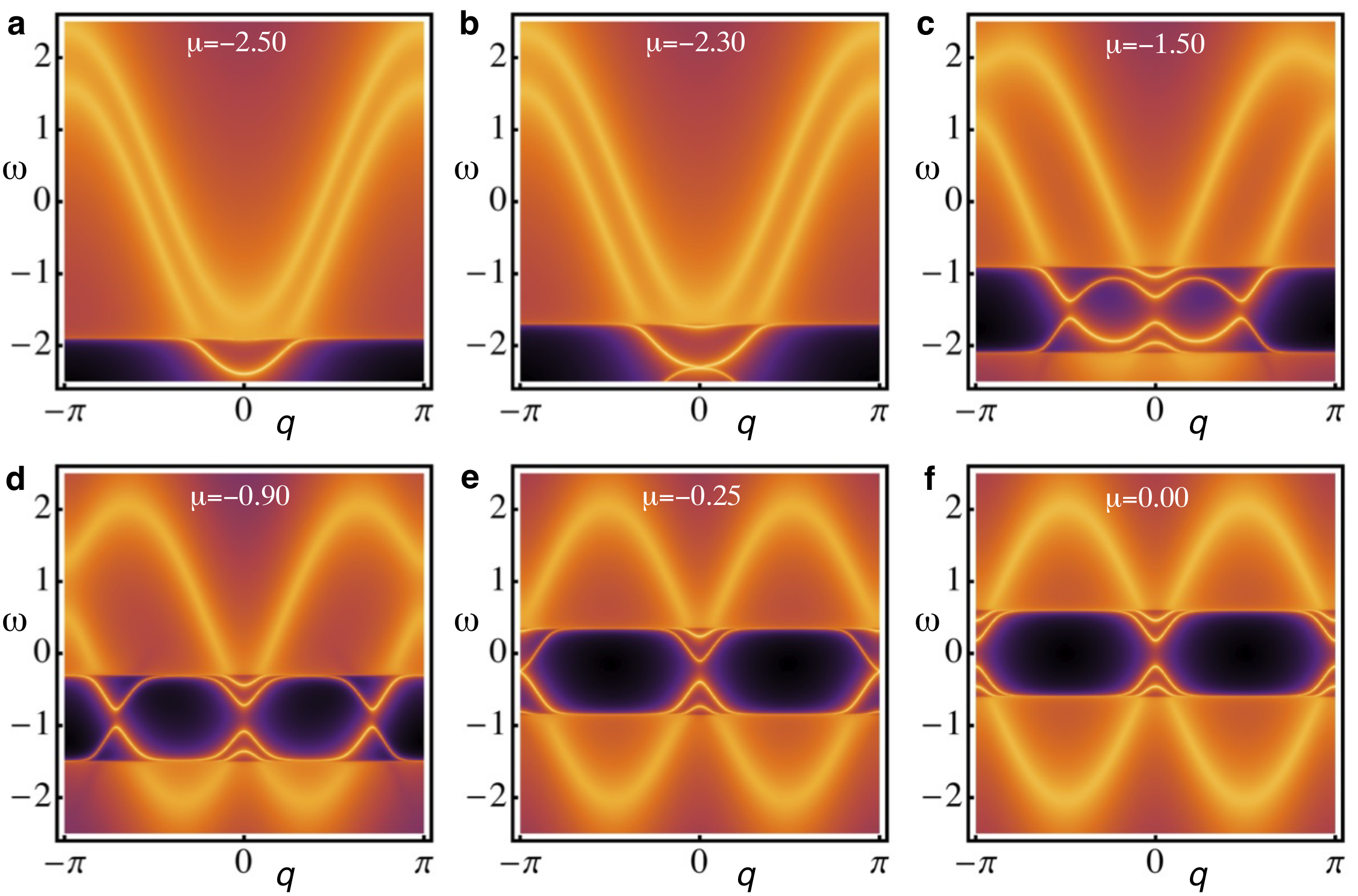}
\caption{{\bf Spectral function of the adatom chain.} $A_{\rm eff}(\omega,q)$ defined in Eq.\ (\ref{hspi14}) is represented as a density plot for several representative values of the chemical potential $\mu$, with the appropriate self-consistently determined spiral pitch $G$ shown in Fig.\ \ref{fig2}a. In panel {\bf a} $\mu$ is below the bottom edge of the chain band  and the system is in the trivial phase. When $\mu$ reaches the band edge a topological phase transition occurs through a gap closing shown in panel {\bf b}. Further increase of $\mu$ puts the system into the topological phase illustrated in {\bf c}, {\bf d} until the gap closes again near half filling {\bf e} placing the system back into the trivial phase {\bf f}. A similar sequence of phases occurs for positive values of $\mu$ for which the spectral function can be obtained by simply flipping the sign of the frequency $\omega$. The dark band around the chemical potential reflects the bulk SC gap.   In all panels frequency $\omega$ is in units of $t$ while $JS=0.4t$, $\Delta_0=0.6t$, $\rho_0r^2=0.05t$ and  $\delta=0.002t$ is used to give a finite width to the spectral peaks.
}\label{fig3}
\end{figure*}
In order to visualize the above effects we display in Fig.\ \ref{fig3} the relevant spectral function, defined as
\begin{equation}\label{hspi14}
A_{\rm eff}(\omega,q)=-{1\over 2\pi}{\rm Im \ Tr} \left[(1+\tau^z){\cG}_{\rm eff}(\omega+i\delta,q)\right]
\end{equation}
where $\delta$ represents a positive infinitesimal. The figure clearly shows how the bands self-consistently adjust to the changing chemical potential as well as the expected topological phase transitions taking place between the trivial and the TSC phases.

Our results thus far relied on the mean field theory and ignored interactions beyond those giving rise to superconductivity. There are several effects that can in principle destabilize the topological state found above but we now argue that the latter remains stable against both interactions and fluctuations. First, one may worry that the adatom magnetic moments would be screened by the Kondo effect at temperatures $T<T_K= \epsilon_F e^{-1/\rho(\epsilon_F) J}$, where $\rho(\epsilon_F)$ is the density of states in the substrate at the Fermi level.  For a normal metal, $T_K$ can indeed be sizeable -- tens of Kelvins -- and the ground state is then non-magnetic \cite{sigrist_rev}.  In the presence of superconductivity, however, $\rho(\epsilon_F)=0$ and a more elaborate treatment of the Kondo problem in the presence of a gap shows that $T_K$ is much reduced \cite{ingersent1}, possibly to zero when $J$ is sufficiently small. Thus, generically, we expect the system to avoid the Kondo fixed point and remain magnetic at most experimentally relevant temperatures.   Electron-electron interactions in the wire are additionally expected to enhance the magnetic gap \cite{loss1,loss2} compared to its non-interacting value $JS$, which will ultimately further improve the stability of the topological phase.

Second, one must consider thermal and quantum fluctuations which will tend to destroy any ordering present in a 1D wire. Since the SC order in the wire is phase-locked to the substrate we may ignore its fluctuations. However, fluctuations in the magnetic spiral order must be considered. In a realistic system spin-orbit coupling will induce a Dzyaloshinsky-Moriya interaction of the form ${\bf D}_{ij}\cdot(\bS_i\times\bS_j)$ in the effective spin Hamiltonian. The latter breaks the SU(2) spin symmetry and pins the spiral order so that the spins rotate in the direction perpendicular to ${\bf D}$. The remaining low-energy modes of such a spiral are magnons.  The relevant spin-wave analysis and the origin of the DM interaction are outlined in the Supplementary material. We find a single linearly dispersing gapless magnon $\omega_q=c|q|$  which will reduce the classical ordered moment according to 
\begin{equation}\label{hspi15}
\langle S^x\rangle \simeq S-a\int_{\rm BZ} {dq\over 2\pi} {1\over e^{\beta\hbar\omega_q}-1}.
\end{equation}
For an infinite wire the integral diverges logarithmically at long wavelengths, signaling the expected loss of the magnetic order in the thermodynamic limit. We are, however, interested in a wire of finite length $L$ where the divergence is cut off at $q\sim \pi/L$. A crude estimate of the transition temperature in this case obtains by assuming $\beta\hbar\omega_q\ll 1$ over the Brillouin zone and setting $\langle S^x\rangle=0$ in Eq.\ (\ref{hspi15}). This gives 
\begin{equation}\label{hspi16}
k_BT^*\approx {\pi S\over \ln{N}}{\hbar c\over a},
\end{equation}
with $N=L/a$ the number of adatoms in the chain. For $N=100$, $S=5/2$ and typical model parameters $t=10$meV, $J=5$meV and $\mu$ appropriate for the topological phase we find $T^*$ of tens of Kelvins (see also the Supplement). Due to the $\ln{N}$ factor $T^*$ is only weakly dependent on the chain length.
 
Our results provide strong support for the notion of self-organized topological state.  Magnetic moments of atoms assembled  into a wire geometry on a superconducting surface are indeed found to self-organize into a topological state under a wide range of experimentally relevant conditions. The emergent Majorana fermions can be probed spectroscopically by the same STM employed in building the structures and will show as zero-bias peaks localized near the wire ends. The system can be tuned out of the topological phase by applying magnetic field $\bB$ which, when strong enough, will destroy the helical order by polarizing the adatom magnetic moments. An attractive feature of this setup is the possibility of  assembling more complex structures, such as T-junctions and wire networks that will aid  future efforts to exchange and braid Majorana Fermions with the goal of probing their non-Abelian statistics \cite{alicea77}.  We also note that the general self-organization principle described above should apply to other 1D structures, most notably quantum wires with nuclear spins considered in Ref.\ \onlinecite{loss1}; however the energy scales are expected to be much smaller due to the inherent weakness of the nuclear magnetism.

{\em Acknowledgement --} The authors are indebted to D. Loss, S. Nadj-Perge, P. Simon, A. Yazdani, and I. Affleck for illuminating discussions and correspondence. The work presented here was supported in part by NSERC and CIfAR.


\newpage
~
\newpage
\vfill
\eject


\section{Supplementary material}

\subsection{Stability analysis of the spiral ground state}

In this section we supply some of the technical details pertaining to the stability analysis of the spiral ground state. In this analysis it is useful to consider a spin-only effective theory that arises upon integrating out the electron degrees of freedom in the relevant Hamiltonian given e.g. by Eq.\ (\ref{hspi1}) in the main text. More generally, we start from a Hamiltonian of the form  
\begin{equation}\label{sup1}
{\cal H}={\cal H}_0
 +J\sum_{i}  \bS_i\cdot(c^\dagger_{i\sigma}{\bm \sigma}_{\sigma\sigma'} c_{i\sigma'})
\end{equation}
where ${\cal H}_0$ describes the electrons in the absence of the exchange coupling to the magnetic moments $\bS_i$.  Assuming translational invariance we can integrate out the electrons to obtain the effective spin-only Hamiltonian 
\begin{equation}\label{sup2}
{\cal H}_S={J^2\over 2}\sum_q\chi^{\alpha\beta}(q)S_q^\alpha S_{-q}^\beta
\end{equation}
where 
\begin{equation}\label{sup3}
\chi^{\alpha\beta}(q)=i\int_0^\infty dt\sum_{i,j}e^{-iq(x_i-x_j)} \langle[s_i^\alpha(t), s_{j}^\beta(0)]\rangle_0
\end{equation}
is the static spin susceptibility evaluated in the ensemble specified by ${\cal H}_0$ and $\bs_i=c^\dagger_{i\sigma}{\bm \sigma}_{\sigma\sigma'} c_{i\sigma'}$. Given the electron Green's function ${\cG}_0(i\omega_n,k)$ the spin susceptibility can be evaluated by analytical continuation of the Matsubara frequency susceptibility
\begin{equation}\label{sup4}
\chi^{\alpha\beta}(i\nu_m,q)={k_BT}\sum_{n,k}{\rm Tr}[{\cG}_0(i\omega_n,k)\sigma^\alpha{\cG}_0(i\omega_n-\nu_m,k-q)\sigma^\beta]
\end{equation}
in the limit $\nu\to 0$. 

In the simplest case of the free noninteracting electrons with ${\cal H}_0=\sum_k(\hbar^2k^2/2m)c^\dagger_{k\sigma}c_{k\sigma}$ the $T=0$ susceptibility becomes $\chi^{\alpha\beta}(q)=\delta^{\alpha\beta}\chi_0(q)$ with the Linhard function
\begin{equation}\label{sup5}
\chi_0(q)={L\over \pi}{2m\over\hbar^2q}\ {\rm ln}\left|{2k_F-q\over 2k_F+q}\right|.
\end{equation}
The latter exhibits peaks at $q=\pm 2k_F$, as already noted in the main text. These peaks constitute the root cause behind the emergent spiral.   

For the SU(2) symmetric model considered thus far the helical ordering can occur in any plane leading to a continuous ground-state degeneracy parametrized by a unit vector $\hat{\bm n}$ perpendicular to that plane. 
One possible mechanism that will break the global SU(2) spin symmetry and select a specific spiral plane is the Dzyaloshinski-Moriya (DM) interaction. It takes the form $\sum_{ij}{\bf D}_{ij}\cdot(\bS_i\times \bS_j)$ and arises in the presence of the spin-orbit coupling (SOC) when spatial inversion symmetry is broken, as is the case in our system of adatoms placed on a substrate. If we assume that the adatom chain lies along the $x$ direction while the surface normal is along $y$ then a Rashba-type SOC of the form $\lambda k\sigma^z$ is permitted in ${\cal H}_0$ with $\lambda$ parametrizing the SOC strength.  DM interaction arises in this situation because as a result of SOC Eq.\ (\ref{sup4}) yields antisymmetric components of the susceptibility $\chi^{\alpha\beta}\sim{\rm Tr}(\sigma^z\sigma^\alpha\sigma^\beta)=2i\epsilon^{3\alpha\beta}$. The D-vector points along the $z$ direction and takes the form ${\bf D}_{ij}=D(\hat{\bm y}\times\br_{ij})$ where $\br_{ij}$ is a vector connecting sites $i$ and $j$. To estimate the magnitude $D$ we continue modeling our wire by the continuum free electron model with Rashba SOC and evaluate Eq.\ (\ref{sup4}) to obtain
\begin{equation}\label{sup6}
\chi^{xy}(q)={1\over 2}[\chi_0(q+Q_\lambda)-\chi_0(q-Q_\lambda)]
\end{equation}
with $Q_\lambda=\lambda m/\hbar^2$ the characteristic SOC wavevector.  At long wavelengths, assuming $Q_\lambda\ll 2k_F$, one can expand
\begin{equation}\label{sup7}
\chi^{xy}(q)\simeq{L\over 12\pi}{q Q_\lambda\over \epsilon_F k_F}.
\end{equation}
This leads to an estimate $D\approx {1\over 12}\Lambda(J/\epsilon_F)^2$ with $\Lambda=\lambda/a$ the SOC strength per adatom site. Taking $J/\epsilon_F\approx 1$ and $\Lambda$ of few meV we obtain $D$ of the order of few degrees Kelvin. We expect the same result to qualitatively describe the tight-binding electrons when the chemical potential lies near the bottom of the band.

The DM interaction discussed above gives rise to a gap $\Delta_{\rm DM}\simeq DS$ towards the out-of-plane spin excitations and for a large adatom spin, e.g.\ $S={5\over 2}$, to a characteristic temperature $T_{\rm DM}\simeq 10$K below which these can be ignored. We emphasize that the DM interaction represents only one possible mechanism by which the spiral plane can be pinned. Another possibility arises from an easy-plane anisotropy which is also allowed by symmetry. 

We now address the effect of the remaining spin-wave fluctuations around the helical ground state pinned to a given plane. These remain gapless as $q\to 0$ and for a chain of a finite length $L$ will destroy the helical order above the crossover temperature $T^*$  given in Eq.\ (13) of the main text. Our goal here is to determine the spin-wave velocity $c$ and estimate $T^*$. To this end we repeat the analysis given in Eqs.\ (1-6) but now including small adatom spin fluctuations $\delta\bS_j$. We thus write
\begin{equation}\label{sup8}
\bS_j=R_j(S\hat{\bm x}+\delta\bS_j)
\end{equation}
where
\begin{equation}\label{sup9}
R_j=
\begin{pmatrix}
\cos{(Gx_j)} & \sin{(Gx_j)} & 0 \\
-\sin{(Gx_j)} & \cos{(Gx_j)} & 0 \\
0 & 0 & 1
\end{pmatrix}
\end{equation}
is the rotation matrix. After the gauge transformation (4) we obtain
\begin{equation}\label{sup1a}
{\cal H}={\cal H}_0
 +J\sum_{j}  \delta\bS_i\cdot(c^\dagger_{j\sigma}{\bm \sigma}_{\sigma\sigma'} c_{j\sigma'})
\end{equation}
with ${\cal H}_0$ given by Eq.\ (5). We now pass to the bosonic magnon variables using the standard Holstein-Primakoff transformation for spin along the $\hat{\bm x}$ direction $S^x_j=S-a^\dagger_ja_j$ and $S^+_j=S^z_j+iS^y_j=\sqrt{2S}(1-a^\dagger_ja_j/2S)^{1/2}a_j$. In the linear spin-wave theory we can neglect terms $O(1/S)$ and the Hamiltonian (\ref{sup1a}) becomes
\begin{equation}\label{sup10}
{\cal H}={\cal H}_0
 +J\sum_{j}\left[-a^\dagger_ja_js_j^x+\sqrt{S/2}(a_js_j^-+a_j^\dagger s_j^+)\right],
\end{equation}
where $s_j^\pm=s^z_j\pm is_j^y$. Once again we may integrate out the electrons to obtain the effective magnon Hamiltonian which takes the following form
\begin{equation}\label{sup11}
{\cal H}_{\rm mag}=\sum_q\left[\Omega_qa^\dagger_qa_q+{1\over 2}\eta_q(a^\dagger_qa^\dagger_{-q}+a_qa_{-q})\right],
\end{equation}
with $\Omega_q=J\langle-s^x\rangle_0+2J^2\chi^{+-}(q)$ and $\eta_q=2J^2\chi^{++}(q)$.
Here $\chi^{\alpha\beta}(q)$ is the static spin susceptibility tensor (\ref{sup3}) evaluated for the Hamiltonian ${\cal H}_0$. The magnon spectrum follows from diagonalizing ${\cal H}_{\rm mag}$ via the Bogoliubov canonical transformation and reads
\begin{equation}\label{sup11a}
\omega_q=\sqrt{\Omega_q^2-\eta_q^2}.
\end{equation}

\begin{figure}[t]
\includegraphics[width = 6.4cm]{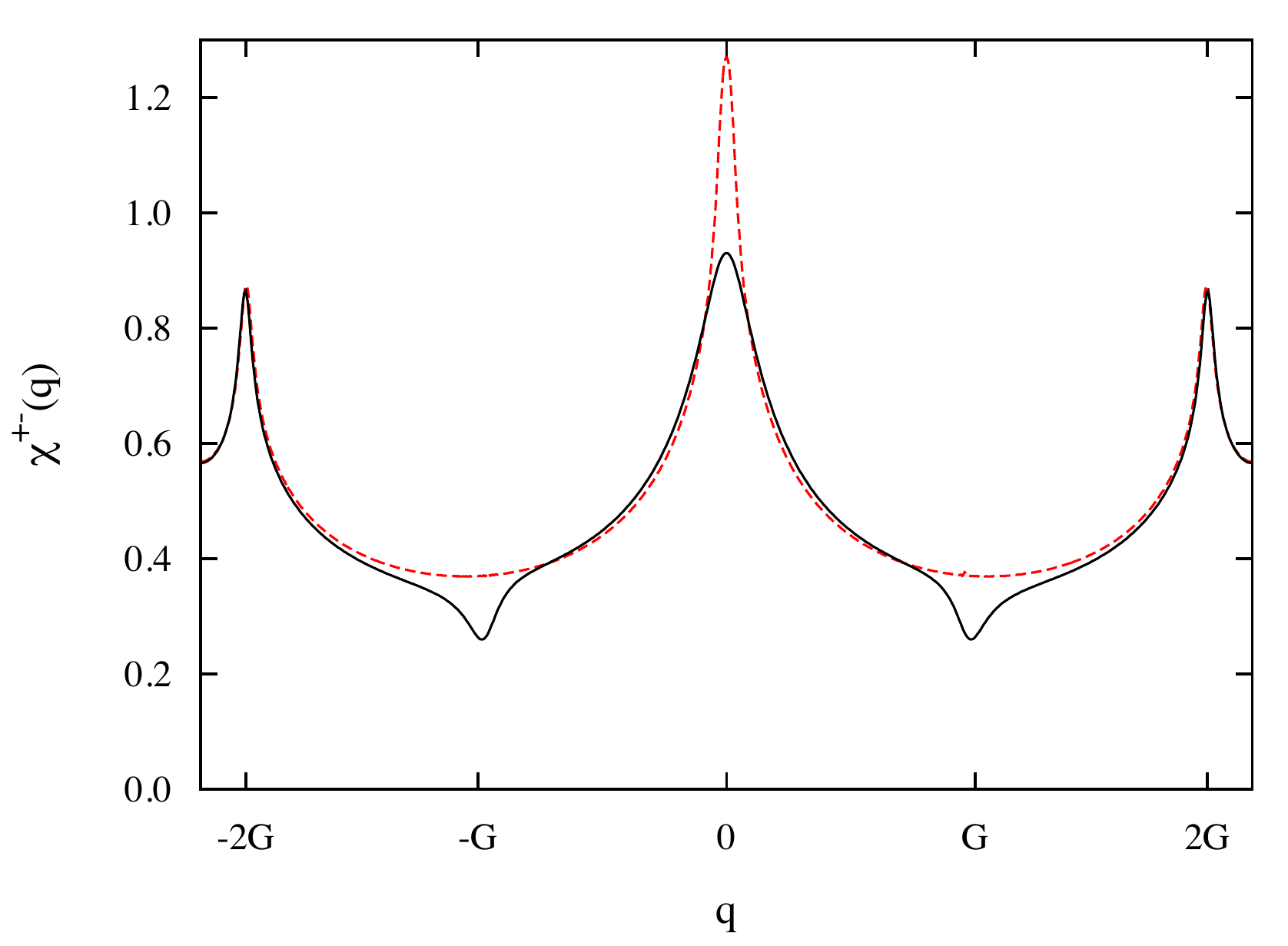}
\caption{Spin susceptibility $\chi^{+-}$(q) in the units of $1/t$ evaluated for the lattice model Eq.\ (5) with $\mu=1.5t$, $\Delta=0$ and $JS=0$ ($JS=0.1t$) for dashed (solid) line. A small temperature $T=0.01t$ has been used to cut off the Linhard divergences. $q$ extends over the first Brillouin zone $(-\pi/a,\pi/a)$.  
}\label{figsup1}
\end{figure}
We have evaluated the relevant components of $\chi^{\alpha\beta}(q)$ for ${\cal H}_0$ given by Eq.\ (5) numerically; some representative results are displayed in Fig.\ \ref{figsup1}. For simplicity and concreteness we focus here on the non-superconducting state. For $J=0$ the susceptibility shows divergences at $q=0,\pm 2G$ inherited from the free-electron Linhard function (\ref{sup5}). For $J>0$ the divergence at $q=0$ is quenched, reflecting the gap that has opened in the electron excitation spectrum at $k=0$ (see Fig.\ 1c). The remaining peaks at $q=\pm 2G$ reflect the fact that a state exists with the opposite helicity (i.e.\ described by a rotation matrix $R_j$ with $G$ replaced by $-G$) which is also a classical ground state of the system. This second ground state, being in a different topological sector, is separated from our chosen ground state by a large energy barrier and will be ignored in our subsequent analysis, which by construction focuses on small fluctuations in the vicinity of a single classical ground state.

To study the long-wavelength magnons we note that it is possible to evaluate the spin susceptibility in the vicinity of $q=0$ analytically. The dominant contribution to the momentum summation in Eq.\ (\ref{sup3}) here comes from the vicinity of $k= 0$ where the Hamiltonian (5) can be well approximated by a 1D Dirac theory with mass $m_D=JS$ and electron velocity $v_F$. To the leading order in $q$ the susceptibility evaluates to $\chi^{\alpha\beta}(q)=\alpha\beta\chi_q$ with $\alpha,\beta=\pm$,
\begin{equation}\label{sup12}
\chi_q={aS\over 4\pi v_F}\left[\ln{\left(1+{\Theta^2\over J^2S^2}\right)}-1-{v_F^2q^2\over 2J^2S^2}\right],
\end{equation}
and $\Theta\sim \epsilon_F$ representing the high-energy cutoff for the Dirac theory. In addition we find $\langle -s^x\rangle_0=4J\chi_0$ as must be the case to satisfy the Goldstone theorem which requires a gapless mode as $q\to 0$. Substituting these results into Eq.\ (\ref{sup11a})  yields the magnon spectrum
\begin{equation}\label{sup13}
\omega_q=4J^2\sqrt{\chi_0(\chi_0-\chi_q)} = c q
\end{equation}
with the spin-wave velocity
\begin{equation}\label{sup14}
c={Ja\over \pi\hbar}K,
\end{equation}
and $K=\sqrt{{1\over 2}[\ln{(1+{\Theta^2/ J^2S^2})}-1]}$ a dimensionless constant close to unity. Taking, as an example, $\Theta/JS=10$ we have $K=1.34$ and for $N=100$ and $S={5\over 2}$ Eq.\ (13) yields $k_BT^*=0.728 J$. For $J=5$meV the crossover temperature $T^*\simeq 36$K. The above analysis suggests that the classical spiral ground state is reasonably stable and the more likely limiting factor in a realistic system will be the substrate superconducting critical temparature $T_c$ which is below $\sim 10$K for most simple metals. 

In closing several remarks are in order. In our calculation of the spin susceptibility we have neglected the SC gap $\Delta$ and worked at $T=0$. Inclusion of the former leads to changes in $\chi^{\alpha\beta}(q)$ near $q=0$ which are small as long as $\Delta<JS$, which is a necessary condition for the topological phase. Such small changes do not significantly alter our conclusions regarding the spiral stability. Similarly, working at non-zero temperature has a negligible effect on the electron spin susceptibility as long as $k_BT\ll JS$ because the electrons are gapped. Since we found both $k_BT_{\rm DM}$ and $k_BT^*$ to be parametrically smaller than $JS$ our analysis is self-consistent in this regard. Finally, one may ask if other spin states might compete with the assumed spiral ground state. We have tried several possibilities, including a spiral with a constant out-of-plane spin component, but all were higher in energy.

\end{document}